\input lanlmac.tex
\input epsf

\ifx\epsfbox\UnDeFiNeD\message{(NO epsf.tex, FIGURES WILL BE IGNORED)}
\def\figin#1{\vskip2in}
\else\message{(FIGURES WILL BE INCLUDED)}\def\figin#1{#1}\fi
\def\tfig#1{{\xdef#1{Fig.\thinspace\the\figno}}
Fig.\thinspace\the\figno \global\advance\figno by1}
%
  


\def\CC {{\cal C}}

\def\CF {{\cal F}}

\def\CL {{\cal L}}

\def\CN {{\cal N}}
\def\CO {{\cal O}}

\def\CR {{\cal R}}
\def\CS {{\cal S}}

\def\CZ {{\cal Z}}
 
 \def\R{\relax{\rm I\kern-.18em R}}
\font\cmss=cmss10 \font\cmsss=cmss10 at 7pt
\def\Z{\relax\ifmmode\mathchoice
{\hbox{\cmss Z\kern-.4em Z}}{\hbox{\cmss Z\kern-.4em Z}}
{\lower.9pt\hbox{\cmsss Z\kern-.4em Z}}
{\lower1.2pt\hbox{\cmsss Z\kern-.4em Z}}\else{\cmss Z\kern-.4em Z}\fi}
\def\p{\partial}

\def\p{\partial}

\font\cmss=cmss10 \font\cmsss=cmss10 at 7pt


\def\IR{\relax{\rm I\kern-.18em R}}
\def\IZ{\relax\ifmmode\mathchoice
{\hbox{\cmss Z\kern-.4em Z}}{\hbox{\cmss Z\kern-.4em Z}}
{\lower.9pt\hbox{\cmsss Z\kern-.4em Z}}
{\lower1.2pt\hbox{\cmsss Z\kern-.4em Z}}\else{\cmss
Z\kern-.4em
Z}\fi}

\def\b{\beta}

\def\pl{{\it  Phys. Lett. }}
\def\prl{{\it  Phys. Rev. Lett. }}
\def\mpl{{\it Mod. Phys.   Lett. }}
\def\np{{\it Nucl. Phys. }}

\def\r{{\rm Re}}
\def\i{{\rm Im}}

\def\hf{{1\over 2}}

\def\sn{{\rm sn} }
\def\cn{{\rm cn} }
\def\dn{{\rm dn} }
\def\z{\zeta }
\def\uy{u_{\infty}}  
 \def\vy{v_{\infty}}

 \def\l{\lambda}
 

\lref\Kon{I. Kostov, \mpl A 4 (1989) 217.}
\lref\GKon{ M. Gaudin and I. Kostov, 
\pl B220 (1989) 200.} 
\lref\KSon{I. Kostov and M. Staudacher, \np B 384 (1992)459.
}
\lref\EZJon{B. Eynard and J. Zinn-Justin, \np B 386 (1992) 558.}
 \lref\EKon{B. Eynard and C. Kristjansen, \np B 455 (1995) 577;
 hep-th/9512052.}
 \lref\DupKos{B. Duplantier and I. Kostov, \np B 340 (1990) 491.}
 
 \lref\Kpo{V. Kazakov, \np B (Proc. Suppl.) 4 (1988) 93;
 I. Kostov, \np B (Proc. Suppl.) 10A (1989) 295.}
\lref\Dpo{J.-M. Daul, preprint hep-th/9502014}
 \lref\BEpo{B. Eynard, G. Bonnet, preprint hep-th/9906130}
 \lref\PZJpo{P. Zinn-Justin,preprint cond-mat/9903385}
  
 
 \lref\bax{R. Baxter, Exactly Solved
 Models in Statistical Mechanics, Academic Press, 1982.}
 \lref\thoo{G. 't Hooft, \np B 72 (1974) 461}
 
 \lref\bipz{E. Br\'ezin, C. Itzykson, G. Parisi and J.-B.
Zuber, {\it Comm. Math. Phys.} 59 (1978) 35.}
 \lref\trico{B. Eynard and C. Kristjansen, \np  B455 (1995) 577.}
 
\lref\mat{V. Kazakov, Phys. Lett. 150B (1985) 282;
F. David, Nucl. Phys. B 257 (1985) 45;
V. Kazakov, I. Kostov and A.A. Migdal, Phys. Lett. 157B (1985),295;
J. Ambjorn, B. Durhuus, and J. Fr\"ohlich, Nucl. Phys. B 257 (1985) 
433}
 
 \lref\bkdsgm{E. Br\'ezin and V. Kazakov, \pl B 236 (1990) 144; M. 
Douglas and S. Shenker, \np B 335
(1990) 635;  D. Gross and A.A. Migdal, \prl 64 (1990) 127; \np B 340 
(1990) 333.}
 
  \lref\bkkm{D. Boulatov, V. Kazakov, I. Kostov, and A.A. Migdal, Nucl. 
Phys. B 275 [FS 17] (1986) 641}
 \lref\Iade{I. Kostov, Nucl. Phys. B 326 (1989) 583}
\lref\Iml{I.K. Kostov, Phys. Lett. B 266 (1991) 42}
 \lref\Icar{I.Kostov, ``Strings embedded in Dynkin diagrams'',  
Lecture given at the Cargese meeting, Saclay preprint
SPhT/90-133}
\lref\Inonr{I. Kostov, \pl 266 (1991) 317.}
\lref\adem{I. Kostov, \pl    297 B (1992) 74.}
\lref\ksks{I. Kostov and M. Staudacher,  \pl B 305 (1993) 43}
\lref\kpz{V. Knizhnik, A. Polyakov and A. Zamolodchikov, Mod. Phys. 
Lett.
A3 (1988) 819.}
\lref\mike{M. Douglas, Phys. Lett. 238B (1990) 176.}
\lref\zinn{P. Di Francesco, P. Ginsparg and J. Zinn-Justin, 2D 
gravity and
 random surfaces, preprint SPhT/93-061,   {\it  Physics 
reports}.}
\lref\Idis{I.K. Kostov,  \np B 376 (1992) 539.}
\lref\Iope{I. Kostov, \pl B 344 (1995) 135.}
\lref\dvv{M. Fukuma, H. Kawai and R. Nakayama, {\it Int. J. Mod. 
Phys.} A 6 (1991) 1385;
R. Dijkgraaf, H. Verlinde and E. Verlinde, \np B 348 (1991) 435.}
\lref\kmmm{S. Kharchev, A. Marshakov, A. Mironov, A. Morozov and 
S. Pakuliak, \np B 404 (1993) 717.}
\lref\bkdsgm{E. Br\'ezin and V. Kazakov, \pl B 236 (1990) 144; M. 
Douglas and S. Shenker, \np B 335
(1990) 635;  D. Gross and A.A. Migdal, \prl 64 (1990) 127; \np B 340 
(1990) 333.}
\lref\kko{V. Kazakov and I. Kostov, \np B 386 (1992) 520.} 
\lref\MFuk{ M. Fukuma and S. Yahikozawa, \pl  B 393 (1997) 316, 
 B 396 (1997) 97, preprint  hep-th/9902169.}

 \lref\df{V.Dotsenko and V. Fateev, Nucl. Phys. B 240 (1984) 312}
   \lref\Nienh{B. Nienhuis, in Phase Transitions and Critical Phenomena,
 Vol. 11, C.C. Domb and J.L. Lebowitz, eds.) Ch. 1 (Academic Press, 1987)
 and references therein.}
\lref\df{V.Dotsenko and V. Fateev, Nucl. Phys. B 240 (1984) 312; 251 
(1985)691}
  
 
  \lref\KM{ V. Kazakov and A. Migdal, ``Recent Progress in the Theory of
Noncritical Strings", \np B 311 (1988) 171.}

 \lref\kleb{I. Klebanov, Lectures delivered at the ICTP 
 Spring School on String Theory and
    Quantum Gravity, Trieste, April 1991, preprint  hep-th/9108019.}
\lref\SDalley{S. Dalley, 
\mpl A7 (1992) 1651.}
\lref\SDalleybis{S. Dalley, \np B 422 (1994) 605.}
 \lref\bkz{D. Boulatov and V. Kazakov, \np B (Proc. Suppl.) 25 A 
 (1992) 38.}
\lref\grkl{D. Gross and I. Klebanov, \np 344 (1990) 475 and
B354 (1991) 459 }
\lref\PGins{P. Ginsparg, 
 Trieste Lectures (July, 1991), hep-th/9112013  .}
   \lref\Hoppe{J. Goldstone, unpublished; 
J.~Hoppe,  "Quantum theory of a massless
relativistic surface ...", MIT PhD Thesis 1982, and  Elementary 
Particle Research Journal
(Kyoto) 80 (1989).} 
\lref\FS{{\it B.A.Fuchs,B.V.Shabat,
Functions of a complex variable},
Jawahar Nagar, Delhi; Hindustan Publ.Corp. 1966
}
\lref\BF{P.  Byrd and M.  Friedman,  {\it Handbook of Elliptic Integrals for 
Engineers and Physicists},  Springer-Verlag, 1954.}
 \lref\AS{M. Abramovitz and I. Stegun,
{\it Handbook of Mathematical Functions}, US Dept of Commerce, 1964.}

  \lref\pzj{ Paul Zinn-Justin,  preprint 
  cond-mat/9909250}
 \lref\kazj{V. Kazakov and P. Zinn-Justin, \np B546 (1999) 647.}
 \lref\HKK{J. Hoppe, V. Kazakov and I. Kostov
 preprint SPhT/t99/072,   LPTENS-99/25, hep-th/9907058, to be published in
 \np B.}
\lref\KKN{V. Kazakov, I. Kostov  and N. Nekrasov,
preprint CERN-TH/98-302,
ITEP-TH-35/98, HUTP-98/A051,  
LPTENS-98/40, and
SPHT-t98/102.}

 \rightline{hep-th/9907060}
\Title{}
{\vbox{\centerline
 {Exact Solution of the Six-Vertex Model }
\centerline{    on a Random Lattice}
 \vskip2pt
}}
%
\centerline{Ivan K. Kostov \footnote{$^{\ast}$}{{\tt kostov@spht.saclay.cea.fr}}
\footnote{$^\dagger$}{member of 
CNRS}}
\centerline{{\it C.E.A. - Saclay, Service de Physique 
Th{\'e}orique }}
 \centerline{{\it 
  F-91191 Gif-Sur-Yvette, France}}

 \vskip 1cm
\baselineskip8pt{
 
\vskip .2in
 
\baselineskip8pt{
\noindent
We solve exactly  the 6-vertex model on a dynamical random lattice, 
using its representation as a large $N$  matrix model. The model 
describes a gas of dense nonintersecting oriented loops coupled to the local 
curvature defects on the lattice. The model can be mapped to the $c=1$ 
string theory, compactified at some length depending on the vertex coupling.
We give explicit expression for the disk amplitude and evaluate the fractal
dimension of its boundary, the average number of loops and the dimensions of
the vortex operators, which vary continuously with the vertex coupling.
}

\bigskip
 \rightline{ SPhT/t99/130}

\Date{November, 1999}  

\baselineskip=16pt plus 2pt minus 2pt


\newsec{Introduction }

 The correspondence between the critical phenomena on flat and random lattices
 has been very useful in various  problems related to random geometry.
 Especially  interesting from this point of view are the $O(n)$
 model\refs{\Kon, \GKon, \KSon, \EZJon, \EKon} and the $q$-state 
 Potts model\refs{ \Kpo, \Dpo,  \BEpo, \PZJpo},
  which allow a neat geometrical 
 interpretation in terms of a gas of  loops or clusters  with
 fugacity depending on the continuous parameter
   $n$ or $q$.  
   Considered on a flat lattice, these statistical models can be mapped onto the 
   6-vertex model \bax , whose infrared behavior is that of a compactified boson. 
   The continuous parameters $n$ and $q$ are then related to the
   compactification radius. The mapping 
    \Nienh , known as Coulomb gas picture, also prescribes to
   install a background  electric charge ``at infinity".

 On an irregular   lattice, the background electric charge, which is 
 in fact  coupled to the local curvature, 
    should be  distributed all over the lattice, which can be achieved only by  
     introducing nonlocal operators at all points with curvature.
   The mapping does not hold any more, which  means that, when considered on 
   a random lattice, the 6-vertex model 
   describes a new type of critical phenomena, in which   the  
   local  curvature of the lattice is involved.  
   
   More concretely, let us   consider the representation \bax\ of the
 6-vertex model  as   a gas  of oriented loops.
     The fugacity of each loop is given by a phase factor, the phase 
      being proportional to the geodesic curvature along the loop.
     On the  flat lattice,   the geodesic curvature is always 
  $\pm 2\pi$ and all loops have the same fugacity.
       On the contrary, on   a lattice with defects,
 the geodesic curvature  can take any value between $\pi/2$ and 
 $\infty$, and the fugacity of the loop depends
 on the local   curvature  of the lattice. 
  In this way, the 6-vertex model on a random lattice allows to 
  study  extended objects  coupled to the
   local curvature, unlike all  previously
   considered  statistical models.

    The 6-vertex model on a random lattice is also interesting 
    from the string theory point of view,
       because   it gives
    a dual definition of the 
compactified $c=1$ string, in which the vortices 
can be constructed explicitly as local operators. 
As such, it was considered first by P. Ginsparg \PGins , 
who formulated the corresponding matrix model.
 The relation between the  6-vertex  model on a random lattice 
 and the $c=1$ string theory 
 has been  further  elucidated by S. Dalley \SDalley, 
who in particular identified the 
compactification radius as a function of the vertex coupling.

In this paper we present the exact solution of the 6v model 
defined on a planar random lattice of any finite size.
The solution is obtained  using  the  equivalence with a special
hermitian matrix model, which is  a generalization 
of the $O(2)$   matrix model \Kon .  
We will calculate the disk amplitude and  investigate its 
scaling behavior.    
 We will evaluate  
 the fractal dimension   of the boundary of the random surface as well as  
    the average  number of loops  per  unit area.   
We will also calculate the dimensions of the $m$-vortex operators $(m\ge 1)$,
which in the loop gas representation 
are the sources of $m$ equally oriented vortex lines.
Our results agree with 
 the  correspondence  between the 6-vertex model on a random surface
 and  the compactified $c=1$  string theory, 
 where   vortices 
represent discontinuities on the world sheet.

The paper is organized as follows. In Sect. 2 we remind the 
definition of the 6v model on a flat and random lattices as well as 
  the  correspondence with  a compactified boson, respectively with 
  the compactified 
  $c=1$ string theory. The model defined on a random lattice 
  depends on two coupling constants: the vertex coupling $\l\in [0,1]$ and the 
  "cosmological constant" 1/T coupled to the number 
  of squares of the lattice. Then we formulate the 
  partition function of the model as a matrix integral, making one 
  step further than the original definition by P. Ginsparg \PGins.
  In Sect. 3 we formulate and solve the saddle point equations for 
  the resolvent of the random matrix for all allowed values of the 
  two couplings.  
  In Sect. 4 we analyze the  the scaling limit of the solution near the 
  critical coupling where the size of the random lattice explodes.
 In Sect. 5 we give a discussion and a summary of the results.   
  The three appendices  contain 
  the proof of the exact correspondence  between the 6v matrix
 model  and the compactified $c=1$ 
string theory (A),   the technical     details of    the calculation (B),
 and   the solution of the $O(2)$ matrix model 
 obtained as the  $\l\to 0$  limit of our general solution (C).

While this manuscript was being prepared, we received  the 
interesting paper by Paul Zinn-Justin \pzj , in which  some of our 
results concerning the critical behavior are derived using
a different technique.

 \newsec{The 6-vertex model on a dynamical random lattice}

 \subsec{Review of the  6-vertex model on a flat square lattice:
 $F$-model, loop gas, vortices, mapping to  a compactified boson.}

 The degrees of freedom of the 6-vertex model on a square lattice
 \bax\  represent  arrows associated with the  links of the lattice.
The admitted configurations of arrows do not contain 
 sources and sinks, so that there is
 an equal number of arrows that begin and end at each vertex. 
 The partition function is defined as the sum
 over all arrow configurations, with Boltzmann factor, which
 is a product of the statistical weights associated with
 the vertices of the lattice.
 On the dual lattice,   the arrows go across the edges and 
 the  vertices are associated with
 the squares.  There are 6 possible vertices, which can be divided 
 into two types depicted in Fig.1, the vertices of each type being 
 related by $\pi/2$-rotations.

 %
\vskip 20pt
\hskip 100pt
\epsfbox{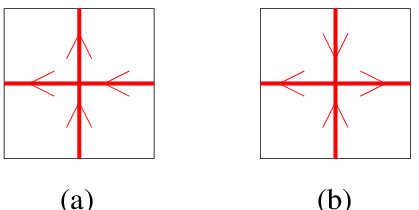}
\vskip 5pt

\centerline{Fig.1: The  two  types of vertices of the
6v model.  }
 
 \vskip  10pt

 \noindent
    For a special value of the spectral parameter the weights of the  vertices 
 of each class have the same Boltzmann weight. 
 This specialization of the 6v model is known as the $F$-model.
  The  weights of the  two distinct vertices (a) and (b) 
  of the F-model  can  be parametrized as
\eqn\weightS{(a)\qquad  a=b =1, \qquad 
(b) \qquad c=  2\cos(\pi\l) }
 where the parameter $\l$ will be assumed to belong  to the interval $[0,1]$.
Following the standard argument of Baxter \bax , we decompose the 
vertices into couples of  line segments making left or right turn as is shown 
in Fig.2. 
The weights \weightS \ are then obtained by assigning a Boltzmann factor
$e^{\pm i\l/2}$ to each line segment, where the sign is positive (negative) if
the line makes a left (right) turn.

 %
\vskip 20pt
\hskip 70pt
\epsfbox{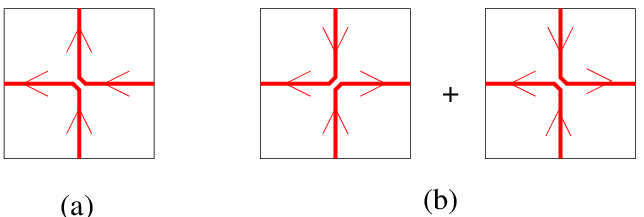}
\vskip 5pt
 
  \centerline{Fig.2:    Decomposing the  vertices as 
 pairs of line segments.}

\vskip 7pt
  
 \noindent
The sum over vertex configurations  on
each surface can be thus 
put in the form of a sum over   fully packed 
closed nonintersecting loops.   
The Boltzmann weight of each loop $\CL$ is the product of all
phase factors along it, 
\eqn\GeoD{ \prod_{{\rm \CL}} e^{\pm{i\pi \over 2}\l}
= e^{i \l \Gamma (\CL) },}
where $\Gamma$ is the geodesic curvature along the loop
\eqn\GAMMa{\Gamma (\CL) = {\pi\over 2} \Big(\#\ {\rm left\ turns}\ 
- \ \#\ {\rm right\ turns}\Big).}
By the Gauss-Bonnet formula,  on the flat lattice   $\Gamma = \pm 4$
and the weight of each  loop (after summing over both orientations) is
$2\cos(2\pi \l)$.
In the 6-vertex  model one can introduce {\it vortex operators} of
  vorticity $m$, representing sources 
 of $m$ equally oriented lines.  
On the flat lattice  $m$ is restricted to be a multiple of 4. 
The 8-vertex model represents a deformation of the 6-vertex
model obtained  by adding   vortices with $m=\pm 4$
with   finite fugacity $d$.

 The  6-vertex model can be rewritten as a solid-on-solid (SOS)  model
  by assigning a  variable $\varphi\in \b\Z$ to  the
  sites of the dual lattice in such a way that the adjacent heights  
  differ by $\pm\b/4$.
  The oriented loops are then  interpreted as    domain walls 
  of height $\pm \b/4$. The $m$-vortex operator creates a discontinuity
  $m \b/4$ at a site. 
  
 At large distances the height variable renormalizes to a 
 free massless boson  compactified on a circle of length $\b=2\pi R$ \Nienh .
  If we choose the scale  so that
the duality transformation acts as
 \eqn\DUALiTY{ R\to {1\over R} \quad {\rm or} \quad \b\to {4\pi^2\over \b} }
 then   the conformal dimensions of the magnetic operators are   
\eqn\DimVor{\Delta_{m}^{\rm vortex} 
=\bar \Delta_{m}^{\rm vortex}= {(m\beta /2\pi)^2\over 4}.} 
   Comparing the singularity of the free energy obtained from the 
   exact solution of the 8-vertex model
 \bax\ with the conformal dimension $\Delta_{4}^{\rm vortex}$, one finds 
 \eqn\Rflat{  \b = \pi \sqrt{1-2\l} .}
The  Kosterlitz-Thouless point is at $\b=\pi$, where the   
lowest vortex operator  ($m=4$)   becomes marginal.

 \subsec{The 6-vertex model on a dynamical random lattice: 
   loops,  
    vortices, 
  mapping to the compactified
 $c=1$ string theory}

 Since the  
weights of the $F$-model   are invariant with respect
  to $\pi/2$-rotations,  the model can be  considered 
  on    any   irregular square lattice $\CS$.  
  The partition function can be again reformulated as a sum over
  loop configurations (Fig.3), each loop $\CL$ weighted by the 
     geodesic curvature \GAMMa
  \eqn\partFUN{ \CF(\CS) = \sum_{^{\rm loop\ conf}_{\rm igurations}} 
  \ \ \prod_{\rm loops\ \CL}
  e^{ i\l \Gamma (\CL) }.}
  In this case   the geodesic curvature \GAMMa\ depends on the 
   geometry of the lattice through the gaussian curvature 
  (the total deficit angle) $\CR(\CL)$ enclosed by the loop $\CL$
 \eqn\GAGAR{\Gamma (\CL)= 2\pi - \CR(\CL).} 
  Therefore $\Gamma(\CL)$ can  be any  integer  
  (positive or negative) times $\pi/2$.

 %
\vskip 50pt
\hskip 50pt
\epsfbox{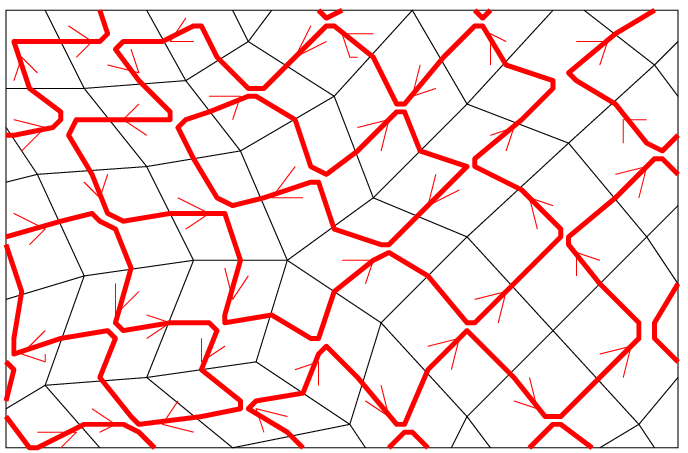}
\vskip 5pt
\centerline{Fig.3:  A piece of the random lattice covered by oriented loops.  }
 
    \bigskip
    \vskip 5pt
 
 We define the partition function $\CF(T,\l,N)$ of the F model on 
 a dynamical random lattice 
  as the average of the partition function 
\partFUN\ in the ensemble of all   connected   
abstract lattices
obtained by gluing squares along their edges. 
 The partition function depends  on 
 the  ``cosmological constant" $T$     coupled to the area ($\#$ squares) 
 $A(\CS)$  and the ``string coupling constant" $1/N$ coupled to
 the genus $g(\CS)$ of the lattice:
 \eqn\PartF{ 
  \CF(T, N,  \l)Ä= \sum_{\CS}ÄN^{2-2g} T^{-A} \CZ(\CS) .}
 It is natural to expect that the critical behavior 
 of this statistical model
  is described by a compactified boson interacting with a Liouville
  field.   
  Indeed, as it was shown in \SDalley,  the  exponential $\CZ=e^\CF$ 
   can be identified with the  
partition function of  a $c=1$ string theory
compactified at  length 
$\beta= 2\pi R$,  related to the vertex coupling as
\eqn\CompR{    R = {1-\l\over 2} \qquad {\rm or}\qquad \beta 
 = \pi(1-\l).}
Here the scale is fixed again by  
 adopting the convention \DUALiTY, 
according to which  the 
self-dual radius is $R=1$ ($\beta = 2\pi$). 
For completeness, we give a short prove of the correspondence  
in Appendix A. 
Note that this is $not$ the same correspondence as on the flat
lattice and there is no reason that it should be the same.

A vortex with vorticity $m$ again represents 
 a source of $m$ equally oriented
lines meeting at a point. However, since the lowest vortex charge is $m=1$,
an $m$-vortex operator creates a discontinuity $m\b$on 
the world sheet. It 
is necessarily accompanied by a conical 
singularity
with curvature $\CR= {\pi\over 4}(4-m)$.  

The   scaling  dimension of 
an
$m$-vortex operator is obtained from the corresponding flat dimension\foot{
 Of course  the length  $\b$  is now given by  \CompR\ and not by \Rflat.} 
 \DimVor\ 
by   the KPZ rule \kpz\ and reads  
\eqn\GRAVD{ \delta_{m}^{\rm vertex} =  {m\beta /2\pi\over 2}= m{1-\l\over 4}.}
The   scaling dimensions of vortex operators with $m=1,2,3,4$ are
always smaller that one, which means that the interval
$0<\b < \pi$ is deep  in the vortex plasma 
  phase.  However, this is not a problem  because 
   in our model the fugacities of all vortices are strictly zero.

 \newsec{The 6v matrix model}    
   
 \subsec{From a complex matrix integral to a Coulomb gas system}

 Ginsparg  noticed   in \PGins\  that   
 the exponential of the partition function \PartF\ can be identified with the 
 perturbative expansion of the following matrix integral  
\eqn\dsixV{ \eqalign{
\CZ_{N}Ä&\equiv e^{\CF(T,N,\l)}= {\int dX dX^{\dag} 
\exp\left(  N\tr[
- \sqrt{T} X X^{\dag} +   X^2 X^{\dag 2}  - \cos\b\
(X^{\dag}X)^2 
  ] \right)\over
\int dX dX^{\dag} e^{-\sqrt{T}N\tr
X X^{\dag}}}
\cr}}
where the integration goes over all $N\times N$ complex matrices $X$. 
    
  At the  point $\b = \pi$ ($\l  = 0$),  this matrix integral  describes 
 the dense phase of the $O(2)$ model on a random triangulation \Kon ,
 or equivalently the compactified $c=1$ string  at the 
 Kosterlitz-Thouless  point  $\beta = \pi$.
 The correspondence with the compactified string for any $\b$ has been
 worked out by S. Dalley \SDalley.  For reader's convenience we give the 
 proof in Appendix A.

 The vortex operators are introduced by adding a source term
 to the matrix potential,
  namely 
  \eqn\SourCE{\CS_{\rm vortices}Ä= N\sum _{{m\ge 1} }Ä {d_{m}Ä\over m}
  (\tr X^{m}Ä+\tr X^{\dag m}Ä).}
The model \dsixV\  in presence of  vertices 
(the random-surface analog of the 8-vertex model)
has been  solved in two
   particular cases: $\l =0,\ d_{1}$ and $d_{2 }Ä \ne 0$ \SDalley\ and
    $c=0, \ d_4 = b$ \kazj .

 \def\11{1\!\! 1}

 The complex matrix integral \dsixV\
  becomes gaussian at the price of  introducing  
  an auxiliary  hermitian matrix $A$:
  \eqn\defLGM{ \eqalign{
\CZ_N  &\sim
\int dA d X^{\dag} dX \  e^{ W(X, A)},\cr
W(X, A)&=N \tr\left(
-  \hf  A^2 +iAXX^{\dag} 
e^{ -i\b/2}-iA X^{\dag}X e^{ i\b/2}   - 
\sqrt{T}X^{\dag} X \right).\cr}}
The result of  the  integration over  $X$  is
a matrix  integral which can be viewed as 
a $\beta \ne \pi$ deformation  of the O(2) 
matrix  model
\eqn\ooD{\CZ_N 
\sim   \int dA\ {e^{-  \hf N\tr  A ^{2}Ä }
   \over  \left|{\det}    \left[\sqrt{T}\ \11 \otimes \11 -i
    e^{ -i \b/2} \  A\otimes \11 +
   ie^{ i \b/2}\   \11 \otimes A)\right]  \right| }.}
The integrand depends only on the eigenvalues $a_1,...,a_N$ of the
 matrix $A$.   After  a liner change of variables, we write the partition 
function as
\eqn\defLGMG{\eqalign{ \CZ_N\sim 
\int \prod_{i=1}^N {da_i\over a_{i}} e^{-NV(a_i)} 
\ \left|\prod_{i\ne j}{ a_i-a_j\over
 e^{ i  \b/2} a_i-e^{ -i  \b/2}a_j }\right|\cr}
}									  
where 
  \eqn\pottt{
  V(a)   = {T\over 2} \left(a^{2}+ { a\over \sin{\b\over 
2} }\right).}

  In the following we will use the more convenient 
  exponential parametrization    
  \eqn\azazaz{a = 
- e^{\phi }}                                             
  in which the   partition function \defLGMG \ 
looks like                                          
   \eqn\defLGMG{ \CZ_N \sim \int 
   _{-\infty}^{\infty}\prod_{i=1}^N d \phi _i \
e^{-N\tilde V( \phi_{i}  )}      
    \prod_{i\ne j}{\sinh 
\hf(\phi_i-\phi_j)\over                                          
     \sinh \hf(\phi_i-\phi_j +i\beta ) 
}}                                
where
\eqn\poTenT{\tilde V(\phi) \equiv  V(-e^{\phi}).}
 Note that when $N$ is finite, the partition functions \defLGMG\ 
and  \defLGMG\ do not
coincide, because the parametrization \azazaz\ assumes that
$a$ is negative. However, the two partition functions  lead to the
same
 $1/N$ espansion of the free energy and the observables.

 \subsec{The saddle point  equations   
 as difference equations for the resolvent}  
 
We are interested in lattices with spherical topology, and therefore by the
   large $N$  limit  of the 
integral \defLGMG.
  In this limit 
  the  spectral density 
  \eqn\RhO{\rho(\phi ) = {1\over N} \sum_{i=1}^{N}\delta(\phi-\phi_{i})}
   can be considered  
    as a classical function, supported by some   real interval 
    $[b,a]$, where $b<a$.
     It is determined by the 
   saddle point equation 
\eqn\sdlptt{ 
 {d \tilde V({\phi}) \over d\phi}   =\int_{ b}^{a}
\! \!  \! \! \! \! \! \! \! \! -  \  d\phi'\ 
  \left({  \rho(\phi') \over \tanh \left({\phi -\phi'\over 2}\right)}-  
   {{1\over 2} \rho(\phi') \over \tanh \left({\phi -\phi'+i\b\over 2 }\right)}
   - {{1\over 2} \rho(\phi') \over 
   \tanh \left({\phi -\phi'-i\b\over 2 }\right)}\right).}
 For a time being we will consider a general potential of the form
 \eqn\PotentiaL{
 \tilde V\phi) = \sum_{n}t_n e^{n\phi}  }
 and later will specify{
 \eqn\ttCST{t_{1 }=- {T\over 2 \sin {\b\over 2}}  , \ \ t_{2}Ä= {T\over 2  }.}
 Following the same logic as in \bipz , we will introduce the complex 
 variable\foot{
   In the context  of the $2D$ string 
   theory, the complex   $z$-plane  represents the target space of the string.
In terms of the $c=1$ noncritical string, the real and imaginary
parts of
$z$   are interpreted correspondingly as the Liouville and target-space directions.  
See, for example, the
argument given in
   \MFuk.}
   \eqn\Zphi{ z = \phi + i t,}
   and formulate the saddle point equation \sdlptt\ as a set of conditions
   on the resolvent 
\eqn\REZ{   W( z) = \hf \int_{ b }^{a} {d\phi  \ 
  \rho(\phi ) \over \tanh {z - \phi\over 2}}\ 
,  }
  which is a holomorphic function on the cylinder $z+2\pi i \equiv z$ cut along the interval
$[b,a]$.
The coefficients of  the  expansion of $W(z)$ at infinity 
\eqn\expMoM{W(z)=\hf +W_1e^{-z}+W_2e^{-2z}+...}
are equal to  the moments of the spectral density
\eqn\MoMEntS{W_n = \int_b^a d\phi \rho(\phi)e^{n\phi} .}

 The spectral density is equal to the discontinuity of the resolvent
  along the cut    
\eqn\fhu{  W(\phi+i0) -  W(\phi- i0) =- 2\pi i  \rho(\phi)   }
and the integral equation \sdlptt\ 
implies the following functional equation  for the resolvent
\eqn\cauSat{   W(\phi +i0)+  W(\phi -i0) -   W( \phi +i \b  )
 -   W(\phi - i\b  )   =
 {d \tilde V({\phi})\over  d\phi}  ,}
where it is assumed that  $\phi  \in [b, a]$.

The  
 normalization condition for the density    can be written also as
 an    integral
 along a contour   $\CC$ surrounding the  interval
 $[b,a]$:
\eqn\normden{1 = \int_{b}^{a}
 d\phi \rho(\phi) =   \oint _{\CC}
 {dz\over 2\pi i}  W(z)  .}
  %
 %
 %

 \subsec{Geometrical meaning of the saddle point equations}
 
Let us   introduce the new potential  
 \eqn\UUU{   U(z) = \sum_{n }{t_{n \ }Äe^{nz}\over 2\sin n{\b\over 2}}}
related to the old one by
 \eqn\UUVV{  \tilde V(z) =   {U(z +i\beta/2) - U(z - i\b/2)\over i} \ ,}
  %
  and consider  the  meromorphic   function 
 \eqn\Jofu{ J(z) = i[  W(z +i\b/2) -W(z -i\b/2)  ]+ U'(z).}
 %
 

The function $J(z)$ is completely determined by the
following four conditions:

 \bigskip
 
 \noindent
(1)-- it  is  periodic with period $2i\pi$:
 $J(z+i\pi) = J(z-i\pi)$;
 
 \noindent
(2)-- it   has two cuts $\{\r z \in [b, a], \i z = \pm i\b/2\}$ in the strip 
$|\i z |\le \pi$,  and 
 satisfies   along the cuts
 a two-term functional equation
 \eqn\fueqJ{J(\phi +i\b/2 \pm i0) = J(\phi-i\b/2 \mp i0), \quad 
 \phi\in [b,a];}

 \noindent
(3)-- its series expansion  at $z\to +\infty$  is
\eqn\Jinfty{ \eqalign{J(z) &= {t_2 \over \sin \b} e^{2z}
+ {t_1 \over 2\sin \b/2} e^z  \cr
&+ 2\sin(\b/2)  W_{1 } \ e^{-z} 
+ 2\sin \b \ W_2 e^{-2z}  +...\cr} }   
 where $W_n$  are  the   moments of the spectral density
 defined by \MoMEntS.

 \noindent
(4)-- it satisfies the normalization condition \normden : 
 if
$\CC_1$ is a contour surrounding the upper cut, then
\eqn\nrmcnd{ \oint_{C_1}{dz\over 2\pi } J(z) = 1.}

 \bigskip

The first two conditions  mean that  the function $J(z)$ is  
a single valued analytic function on 
the torus
obtained from the  strip $|\i z | \le \pi$ by cutting it  
 along the intervals
 $\{\r z \in [b, a], \i z = \pm i\b/2\}$, then identifying its two edges
 ($\i z = \pm \pi$) as well as  
the  $_{ \rm lower} ^{\rm upper}$ edge of the first  cut  
with the  $_{ \rm upper} ^{\rm lower}$  edge of the other one. 
 The two main cycles of this  torus  are homotopic to the contour 
 $\CC_{1}$ surrounding  the upper cut, and the contour
 $\CC_{2}$ connecting the two cuts.  

  The variable $J$ is periodic along both cycles while 
  the variable $z$ is periodic  along the cycle $\CC_1$  and has discontinuity
   $i\b$ along the cycle
  $\CC_2$.
 Therefore, if we find  a canonical parametrization  
  \eqn\Zzu{ z = z(u), \ J= \zeta(u)}
  for which
 the contours  $\CC_{1}$ and  
 $\CC_{2}$ are the images of the   periods 
  $\omega_1$ and $\omega_2$ of a flat rectangle,
  then   the functions $z(u)$  and $  \zeta(u)$ will 
 satisfy  the following periodicity  conditions
  \eqn\PERiody{ \eqalign{ 
  z(u+  \omega_1 ) &= z(u), \ \ \ \  z(u+  \omega_2 )= z + i\b,\cr
  \z (u+  \omega_1 ) &= \z(u), \ \ \  \z (u+  \omega_2 ) = \z(u).
  \cr}}

Therefore, once the elliptic parametrization 
    is found, the problem is essentially resolved\foot{Our method is 
    a natural generalization   of the method  applied by 
    J. Hoppe, V. Kazakov, the author and N. Nekrasov to resolve 
     technically similar problems \refs{\KKN ,\HKK}, and which    was 
      originally proposed by J.  Hoppe 
    \Hoppe. }. 
   The parameters of the solution  are  fixed by 
   the normalization condition for the spectral density and the  
   behavior of the function $J(z)$ at $z\to\infty$, which is determined 
   by the
   potential $U(z)$.  
    
    Let $\uy$ be the point in the periodic rectangle corresponding
    to $z\to\infty$. The map is assumed to be regular, which means that 
     $e^{-z(u)}$ can be Taylor expanded at $u=\uy$. 
      Therefore    
 $e^{z(u)}\sim (u-\uy)^{-1}$ and  the  third condition \Jinfty\  
 says that the function 
   $\z(u)$ has a second-order pole at $u= \uy$ and
   fixes  the singular part  of its Laurent expansion in $u-\uy$.

 \subsec{ Explicit solution in the elliptic parametrization}

 We will identify  the two periods of the torus with 
 the standard  quarter-periods 
 of the Jacobi elliptic functions
$\omega_1 = 2K$ and $\omega_2 = 2iK'.$ 
The  $z$-cylinder with the two cuts is parametrized by the
   rectangle  
\eqn\RECT{ |\r u |\le K,   |\i u|\le K'.}

The  function  $\z(z)$ is periodic with periods $2K$ and $2iK'$ and
its only singularity in the rectangle \RECT\ is a double pole at 
  $u=\uy$. The residue of this pole is zero, since otherwise there 
would be another pole somewhere in the rectangle.  
  The generic function with such properties    
(the Weierstrass elliptic function) depends on two constants $A$ and $B$
 \eqn\SOLUTIONZ{\eqalign{\cr
 \z(u)&=  A + B {1\over \sn^2 (u- \uy)}
.\cr}}
The function $ z(u)- {a+b\over 2}$ is antisymmetric in $u$, which 
  means that $e^{z(u)}$ has one simple pole at  $u=\uy$ and 
  one simple zero at $u=-\uy$. 
  It is therefore natural to look for it    as a ratio of
  Jacobi theta functions. One can easily  
  check  that the   function
\eqn\SOLUTIONz{\eqalign{
 e^{z(u) }&= \ \  e^{a+b\over 2}  {H(  \uy +u) \over H( \uy -u)} \cr}}  
   satisfies   the periodicity condition
   \PERiody\  if the parameter $\uy$ is related to the period $2K$ by
   \eqn\uinfty{ \uy = {2\pi - \b\over 2\pi} K = {1+\l\over 2}K.}  
   The  function \SOLUTIONz\ gives the desired regular map 
    between the strip with two cuts in the $z$ plane and 
    the  rectangle \RECT , as is shown  in Fig. 4.

  The  two horizontal edges of the 
$u$-rectangle  are mapped to   the two cuts in the $z$-strip. 
Its  two vertical edges   are mapped to the line made of the segments
$[i{\b\over 2}, i\pi]$ and $ [- i\pi, -i{\b\over 2}]$.  
The function $z(u)$ along the 
$^{\rm upper}_{\rm lower} $  cut  is given by
\eqn\ZONCUTS{ z (\pm iK' + y ) = {a+b\over 2}  \pm i{\b\over 2}
 + \ln { \Theta ( \uy +y) \over \Theta ( \uy 
-y)}.}
The endpoints   of the cuts are the    four 
 branch points of  the inverse map $u=u(z)$, determined by
the condition $ dz/du = 0$. The symmetry of the map
\SOLUTIONz\ allows to determine  the four  branch points through a single 
 parameter $u_{b}\in[0, K]$:
\eqn\BRPTS{ a\pm i {\b/ 2} = z(u_b \pm iK')  ,\ \ \ 
b\pm i {\b/ 2}= z(-u_b \pm iK')  .} 
The parameter $u_{b}$ is determined from 
 \eqn\BRCHP{ Z(\uy + u_{b}) + Z(\uy - u_{b})=0,}
where $Z(u) = {\Theta'(u)\over \Theta(u)}= E(u) - {E\over K} u$ 
 is the Jacobi Zeta function \BF.
   In particular, if $\beta = \pi$, then 
 $\uy = u_{b}$.
Finally,  the length of the cut is given, according to \BRPTS, by
\eqn\AMINUSB{  a-b = \ln  { \Theta ( \uy +u_{b} ) \over \Theta ( \uy 
-u_{b})}.}

Note  the following symmetry, of the solution, which reflects 
the symmetry $\b  \to 2\pi - \b$ (or $\l\to -\l$) of the
original matrix 
integral
$$\eqalign{
\b &\to 2\pi - \b, \cr z(u) &\to - z(K-u), \cr   \z(u) &\to - \z(K-u).\cr}$$

%
\vskip 20pt
\hskip 20pt
 \epsfbox{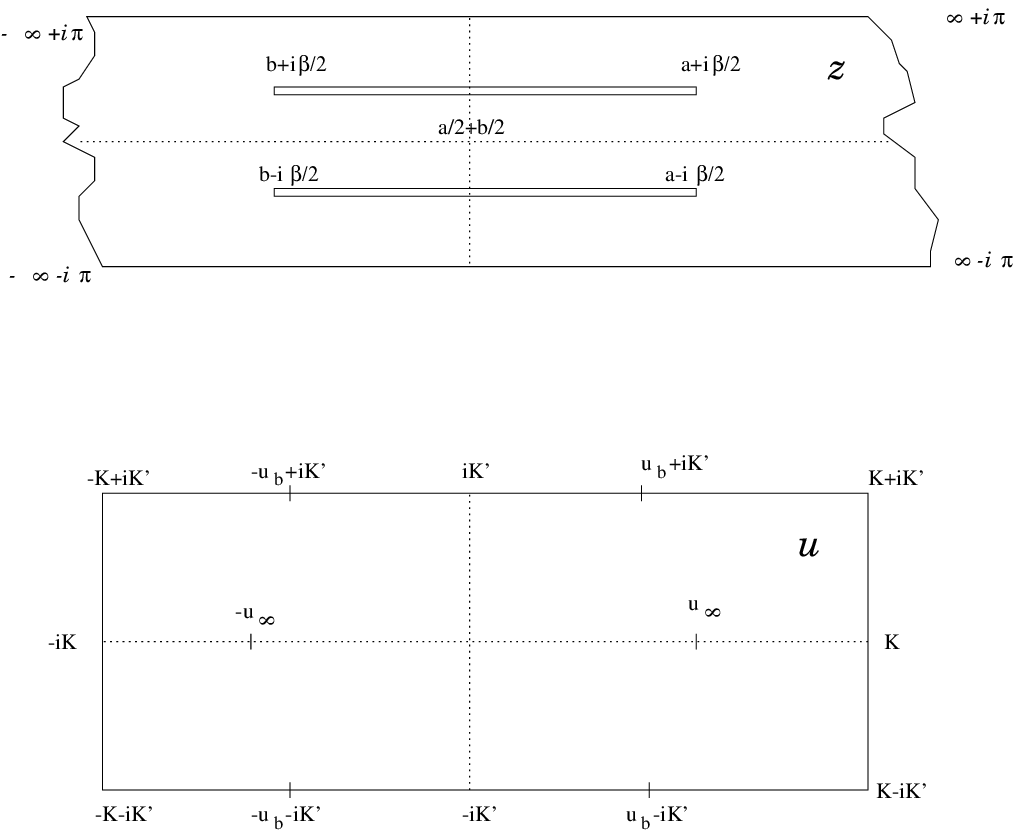}
\vskip 5pt
\centerline{Fig.4:  The domains of the variables $z$ and $u$.  }

\noindent
Now we have to fix, using the asymptotics 
\Jinfty\ and the normalization condition \nrmcnd , the 
 four parameters of the solution, namely
 
\noindent
-- the 
coordinate of the middle point of the cut ${a+b\over 2}$, 

\noindent
--
the coefficients $A$ and $B$,  

\noindent
-- the elliptic nome $k^{2}$, which 
determines the two periods $K$ and $K'$.

 The asymptotics at infinity gives three independent conditions, which
 together with the normalization determine completely the
 four parameters.
  A relatively simple  calculation  (see  Appendix B) gives 
  \eqn\POWERONE{ e^{a+b\over 2} 
 =  { \cot{\b\over 2}\over 2}  \ { H'(0)  \over
    H'(2\uy )},}
 \eqn\CSTB{ B =  {\pi  \l  T\over T^*}   
 \ {H^2(2\uy)\over H'^2(2\uy)},}
 %
 %
  
  \eqn\modULUSb{
  {T^{*}Ä\over \l T}=
  {(K-E)    [Z(2 \uy) +  \cn\dn(2 \uy)]- K {Z(2 \uy) 
 \over \sn(2 \uy)}  
  \over  \left[Z(2 \uy) 
+ {\cn\dn\over \sn}  (2 \uy)
\right]^2}.}
   Here we introduced the  critical  value of $T$
 \eqn\CouplTC{   T^{* }
 =
 16  \pi \l  \  { \sin^{3}{\b\over
 2}   \over   \cos {\b\over
 2}} }
  for which      $K'/K\to 0$.
   The latter coincides with  the
 critical coupling $b^*=1/T^*$  obtained by  P. Zinn-Justin \pzj .


 \newsec{The scaling limit}

 \subsec{Explicit form of the solution in the scaling limit}

 \def\vy{v_{\infty} }

 We would like to explore the scaling behavior 
 near the critical point $T\to T^*$,  where $e^{b}\to 0$,
  which in our parametrization
  corresponds to  the limit  $K\to\infty, 
 K'\to \pi/2$.  
    In this limit the solution can be expanded in the
dual modular parameter
\eqn\QqQ{ \eqalign{ q &= e^{-\pi K/K'}. \cr}}
The expansion takes simpler form if we  rescale slightly the variable $u$ to 
\eqn\vVvV{ v = {\pi\over 2 K'}     u  
}
which belongs to the domain $|\i v |\le {\pi\over 2}, |\r v| \le  
\hf \ln{1\over q}$.
 %
 %
 
 Our  solution \SOLUTIONZ --\SOLUTIONz\ expands as a series in $q$  as 
 follows:
  \eqn\zETAE{\eqalign{
  \z(u)
  &= A-B{ E'\over K'}-      B { \pi^2 \over 4 K'^2} \left[
   {1\over \sinh^2 (v-\vy)}
   + 
  \sum_{n=1}^{\infty}8n \ { q^{2n} \cosh 2n(v-\vy)\over 1- q^{2n}}
  \right]\cr}}
\eqn\ZedUU{\eqalign{
z(v) 
 & = \ln {a+b\over 2}  -(1+\l)  v + 
 \ln {\sinh( \vy +v)\over \sinh( \vy-v)}
+4 \sum_{n=1}^{\infty}  {q^{2n}\over n}
{\sinh 2n\vy \sinh 2n v\over 1-q^{2n}}, \cr
}
}  
where $$\vy =    
 {1+\l\over 2} \ln {1\over q}.$$ 
    

 In order to find the    $q \to 0$ asymptotics of \zETAE \ and
 \ZedUU , we need   the explicit expressions for the parameters
 $T, a, b, B$  as functions of $q$.
  In the following we will use the symbol ``$\approx$"  for 
 ``equal up to subleading terms in $q$".
  %
   From
  \AMINUSB
  ,
  \POWERONE
  ,
  \CSTB
  \ and 
  \modULUSb
  \ we get
  %
 \eqn\parameterss{\eqalign{ \Lambda \equiv   {T-T^{*}Ä\over T^{*}}Ä
 & \approx   {1-\l\over \l}
 q^{1- \l } \ln {1\over q} \cr
 e^{a+b\over 2} \ &\approx    {\tan{\pi\l\over 2}\over 4\l}
     q^{{1-\l^{2}\over 4}}    \cr
   e^{a-b\over 2 }\  &
\approx  {(1+\l)^{1+\l\over 2}
(1-\l)^{1-\l\over 2}\over 2 } \ q^{-{ 1-\l^2 \over 4}}\cr
    B \ \ &
\approx {T\over T^*}\ {\pi\l\over (\l +q^{1-\l})^2}\cr }}

  We see that indeed the position of the left cut  tends to $-\infty$
  \eqn\eftCUT{ e^{ b  } \sim 
  q^{{ 1-\l^2 \over 2}} \to 0,}
while the right cut   has a finite limit at $q \to 0$
 \eqn\RHSofCUT{ e^{a  }\approx   { \tan{\pi\l\over 2
 }\over 8\l}    \ 
{{(1+\l)^{1+\l\over 2}
   (1-\l)^{1-\l\over 2}}}
. }

\bigskip

Now we are ready to determine the scaling behavior of the solution 
 \zETAE-\ZedUU.  
The interesting scaling  behavior   
  is associated with the vicinity
of the  left   pair of   branch points of  $J(z) $, that is
at 
$$b\pm i {\pi\over 2}  =z\left( -v_b \pm i{\pi\over 2} \right).$$
It is therefore convenient 
to introduce  a new parameter $\tau$   
\eqn\uTau{  \tau= v+  v_{b}  ,   }
  such that the    
  two left branch points occur  at $\tau = \pm {\pi\over 2}$.
 From \BRCHP\ we get 
\eqn\BTCHPT{v_{b}Ä \approx  \vy -\hf \ln{1+\l\over 1-\l}.}
    
  and 
  the point $+\infty$ of the  $z$-plane corresponds    to
 $$\tau_{\infty} \approx 2\vy - \hf \ln  {1+\l\over 1-\l } .$$
 Assuming  that $ |\tau  |\ll \tau_{\infty} $
 and
 neglecting the  subleading terms we get
 \eqn\tRLzt{
 \zeta (\tau ) \approx    {4\pi\over \l}
 \left(q^{1+\l}Ä {1-\l\over 1+\l} \ e^{-2\tau}Ä
 +q^{1-\l}Ä {1+\l\over 1-\l} \ e^{2\tau}Ä  \right)}

 \eqn\tRzt{ e^{ z(\tau )} \approx   C \ 
   \left((1+\l)\ e^{(1-\l)\tau}Ä- (1-\l)\ e^{-(1+\l)\tau}Ä\right)}
  where 
 \eqn\cstC{ C \approx
    { \tan(\pi\l/2) \over 4\l }{ q^{1-\l ^2\over 2}Ä\over
       (1+\l)^{1-\l\over 2}Ä(1-\l)^{1+\l\over 4}}.}
  

   \subsec{The scaling behavior of the spectral density.  
   Critical exponents associated with the entropy of the loops }
   
    Let us first check that the ``string susceptibility" exponent
    $\gamma_{\rm str}$ is   zero for all $\l$.  
       Consider   the first momentum of the spectral density,
 whose singular behavior is that of a sphere with one puncture
 $$W_1\sim {\p \ln \CZ \over \p \Lambda} \sim {\rm constant} \ + ({\rm another \ constant})\ \Lambda^{1-\gamma_{\rm
 str}}.$$
The singular behavior is   that of the coefficient  $B$,
which is given by the third equation \parameterss:\ 
$B- B^* \sim  q^{1-\l}   $.  
On the other hand, by the first equation \parameterss
\eqn\QULAM{  {q^{1-\l}\over \l} \approx    {\Lambda\over  \ln {1\over 
\Lambda}}}
which implies  that  indeed $\gamma_{\rm str}=0 $ and the   free energy 
has logarithmic singularity 
$$ \ln \CZ \sim { \Lambda^2\over  \ln {1\over  \Lambda}}.$$

 Then we look for the scaling behavior of the spectral density and the
 resolvent.     The spectral density is related to the function $\z(v)$ 
      by
       \eqn\DENSity{\rho(z) =
  {  \zeta(v+i\pi/2)   +   \zeta(u-i\pi/2)  \over 2\pi} 
  }
       and the resolvent \REZ\ is obtained by inverting the relation \Jofu
    \eqn\zW{  
   W (i\pi +\tau ) = W (i\pi -\tau ) 
   =  {\tau \over \pi} \left[ \z\left( \tau + i{ \pi\over 2}\right) - 
   \z\left( -\tau - i{ \pi\over 2} \right) \right] .}
 It is easy to see that the  function  $W(z)$  
 has only one cut on  the physical sheet (the strip
 $|\i z| \le \pi$),
  which goes  along the interval   $z\ge b$.  
   On the other sheets there is an infinite 
  number of cuts (at least for irrational $\l$), placed at
  $\i z =  i\pi n \pm \l, \ n\in \Z.$

 At distances $  e^{z}\gg e^{b}$, the   relation  
 between $z$ and $\z$ has
  two branches  
 \eqn\TwoChann{ \zeta_\pm (z) \sim q^{1\pm \l\over 2} \exp\left({2z\over 1\mp
 \l}\right) }
  associated with the asymptotics at $ e^{\tau}\  {^{\gg}_{\ll} }\ 1   $.
   It is the smaller exponent which is the relevant one, 
   but in the limit 
  $\l\to 0$ the two exponents coalesce and   produce the  
 observed logarithmic behavior  of the spectral density
  the $O(2)$ model. Correspondingly, the    $\tau$-factor  in 
    the resolvent \zW\ becomes $\tau^2$ in this limit(see  eq. (3.41) 
    of \Idis ).  The limit $\l\to 0$ of the solution is worked out
    in Appendix C.

 Now let us find the scaling of an important observable: 
 the typical length $L$ of a   
 loop  on the world sheet.  
 It is characterized by the critical exponent $\nu$ \DupKos
 \eqn\LALA{ L\sim A^{1\over 2\nu}.}
 A related quantity is the average number $\CN$ of loops  on a 
 worldsheet of area $A$. Since $A$ is the total length of all loops,
 \eqn\NANANA{\CN = {A\over L}\sim A^{2\nu -1\over 2\nu}.}
To  follow  the analysis of \DupKos\ and
\Idis,  it is convenient to return to the original variable
 $a = e^\phi$ and consider the density
 \eqn\spds{\rho(a) = e^{-\phi} \rho(\phi)  \sim a^{1-\l\over 1+\l}}
 and the resolvent $W(w)= \langle \Tr {1\over w - A}\rangle $
 as a function of   the complex variable $w=-e^z$ 
 \eqn\resAA{W(w) = -e^{-z} W(z) \sim    w^{1-\l\over 1+\l}\ln w.}
The resolvent $W(w) $ has a cut along the interval
 $-\infty <   w<-M$, where
 \eqn\InvCorrLength{M=e^b\sim q^{1-\l^2\over 2}\sim 
\left(\Lambda\over \ln {1\over  \Lambda}\right)
^{1+\l\over 2}}
  gives the scale for the length of the boundary.   
  Since $A\sim \Lambda$ and $L\sim 1/M$ (for  detailed arguments 
  see \Idis), we conclude that
\eqn\NUNUNU{\nu = { 1\over 1+\l}.}   
  and the number of loops grows with the area of the world sheet
  as
  \eqn\NANNu{\CN \sim A^{1-\l\over 2}.}
 Let us also mention that the exponent $\nu$  determines also the 
 fractal dimension of the
  e boundary of the disk amplitude, which is equal to
  ${1\over 2\nu}$.

 Finally, let us calculate the dimensions of the $m$-vortex operators
 representing sources of $m$ equally oriented lines.
 The correlation function of two such operators is 
 obtained by calculating the "watermelon" configuration
 made of $m$ nonintersecting lines relating the two points,
as the one shown in Fig.5. 
 The singular behavior of such an  amplitude is related to the
 scaling dimension $\delta_{m}^{\rm vertex} $
 of the $m$-vortex operator as 
 \eqn\amPPmmm{\chi_m \sim \Lambda^{2\delta_m - \gamma_{\rm
 str}}.}

%
\vskip 20pt
\hskip 55pt
 \epsfbox{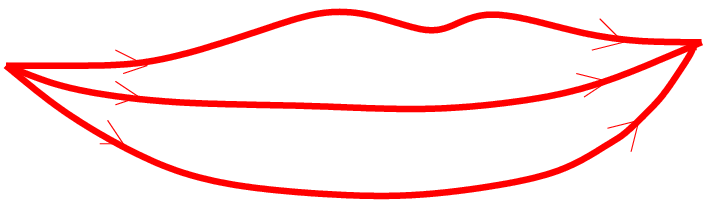}
\vskip 20pt
\centerline{Fig.5: A line configuration for the correlator of
two 3-vortices.    }

\bigskip

 Proceeding in the same way as in Sect.4 of ref.\DupKos , 
 one can extract the 
 scaling dimension  of the $m$-vortex operator
    from the scaling behavior of the spectral density
 $\rho(a)$ (eq. \spds). The  amplitude corresponding to a
 configuration with 
 $m$ open lines is given (up to the phase factors
$e^{i\l\Gamma}$ associated
with the $m$ lines, which can be neglected)   by
the convolutive integral in  eq. (4.28) of ref. \DupKos, and has
the  following singular behavior
 \eqn\CHII{\chi_m \sim M^{{1+\l\over 1-\l}m} \sim \Lambda^{{1-\l\over 2}m}.}
Compared it with  \amPPmmm , we find the  same dimensions  \GRAVD  
 \eqn\GRAVDb{ \delta_{m}^{\rm vertex}  = m{1-\l\over
 4},}
which confirms the interpretation of the $m$-vortex operator  as
a  discontinuity $m\b$ on the world sheet.
 
\newsec{Summary and discussion}

In this paper we  gave   the exact solution of the 6-vertex model on a 
dynamical random lattice with spherical topology.   
  The model describes a gas of densely packed
loops whose weights are coupled to the local curvature of the 
  lattice.   
We have calculated  the  partition function of 
 the disk  with given length $L$ and area $A$.
  For any value of the vertex coupling 
 $\l$,  the critical singularity of the free energy 
   corresponds to a theory with 
 central charge $c=1$, but the fractal dimension of the boundary
 of the disk   changes continuously  from $1/2$   to $1$.
 The limiting value is achieved at $ \l = 1$, where
 the boundary has the same dimension as the bulk.
 This reminds what happens in the dense phase of the $O(n)$ model,
 where at $n\to 0$ the world sheet is all eaten by the boundary.
  Indeed, the exponent ${1-\l\over 2}$ for the growth of the number of loops 
   vanishes when  $\l\to 1$, which means that in this limit the 
   effective fugacity 
  of the loops renormalizes to zero due to the     fluctuations 
  of the curvature.
 We also calculated the scaling dimensions of the $m$-vortex operators,   
representing sources of  $m$ equally oriented 
open lines on the world sheet.

All our results agree with the interpretation of the 6-vertex model
as a  special realization  of the compactified $c=1$ string theory. 
From the  string theory point of view  the model is interesting 
 with the possibility to treat very explicitly the vortex operators,
 which are difficult to construct in the standard  realizations of the $c=1$ 
  string. 
 
The 6-vertex matrix model describes the $c=1$ string theory compactified
at length  $\b <\pi$, which is less than  1/4 of the Kosterlitz-Thouless
  length $\b=4\pi$, associated with the most relevant vortex 
  operator ($m=1$). 
  In order to describe a theory compactified at length up to $p\pi$,
  it is sufficient to consider a chain of $p$ coupled matrices.
   The diagonalization
  can be performed in a similar way and one obtains a chain of integral
  equations of the type \sdlptt , which  seems to be  exactly  solvable
  as well.
    
  The loop gas formulation of the $c=1$ string theory places a 
  bridge between the  description with continuous target space  {\it via}
  matrix quantum mechanics \ref\KazMig{V. Kazakov and A.A.  Migdal,
  \np B311 (1988) 171} and the matrix models \adem\
  for the strings with discrete target space 
  (the ADE models on a random lattice) \Iade,
  which describe all minimal conformal theories coupled to $2d$ gravity.
 In particular, it is clear why the   
    string diagram technique based on a world sheet surgery
       \Idis \
    works only for strings with discrete target space. 
   A discrete target space without loops, i.e., a Dynkin graph of 
   ADE type, can (and should) be immersed in the continuum
   by placing  its adjacent points at distance $\b=\pi$, 
   for which the curvature 
   defects do not produce phase factors for the domain walls.
       In the case of the $c=1$ string compactified at length $\b$,
       this diagram technique    works only at the point 
       $\b = 1$ (the $O(2)$ model).

  The nonperturbative equivalence with the $c=1$ string theory 
  means that the 6v matrix model is exactly solvable in the so called
  double scaling limit \bkdsgm.
  The integrability of this model follows also 
    directly   from the fact that the partition function
     \defLGMG\ is a $\tau$-function of the KP hierarchy.
     Using the scaling properties of the partition function integral, 
     one can 
convert the 
KP equation into  an ordinary differential equation, as it has been 
done in ref. \KKN.

\newsec{Acknowledgments}
%
\noindent The author is  grateful to J. Hoppe  and 
V. Kazakov,  
who participated  in  an early   project related to this work, 
and to P. Zinn-Justin and J.-B. Zuber for   critical reading of the manuscript.
This research
is supported in part by European  TMR contract ERBFMRXCT960012 and EC 
Contract FMRX-CT96-0012.


\appendix{A}{Equivalence to the $c=1$ compactified string }

 Let us integrate with respect to the angular variables in the 
 original matrix integral \dsixV. We decompose 
 $$XX^{\dag}Ä=  U  {x}U ^{-1}$$
 with  $ U \in SU(N)$ and $x={\rm diag}(x_{1}, \ldots, x_{N})$,
 and  perform  the $U(N)$ integration with the help of
 the Harish Chandra-Itzykson-Zuber
 formula. Using the Cauchy identity, the  integrand can be written as a
 determinant, and the partition function takes the form 
\eqn\invosc{\CZ_N =  
\int_0^{\infty}   dx_1 \ldots dx_{N }  \ \det_{ij} K(x_i, x_j)}
with 
\eqn\kerNel{ K(x,x') = \exp   \Big(-\hf N \sqrt{T} (x+x')
 +   Nxx' - \hf N (x^2 +x'^2) 
\cos \b  \Big).}
We get rid of the linear term by  a linear change in  the integration variable
\eqn\ylam{   y=   \sqrt{N \sin\b} \, x - y_ 0,
\qquad
y_0=    {\sqrt{N T \sin\b} \over \sin^2{\b\over 2}} 
= 2\sqrt{2NT/T^*},  
 }
after which the kernel becomes  identical (up to normalization)
to that of the inverse oscillator 
\eqn\operK{ K(y,y')= {1\over \sqrt{2\pi \sin \b}}\ e^{
- { 2yy' - \cos \b  (y^2 +y'^2) \over 2\sin \b}}=
\langle y| e^{ \b (\hf \p_y^2 +\hf y^2)}|y'\rangle ,
}
and the
integration is to be done in the semi-infinite interval 
$$- y_{0}  <y< \infty.$$  
Hence the  partition function \invosc\ describes a system of
$N$ nonrelativistic fermions  in 
inverse oscillator potential, stabilized by  a 
  wall at distance $  y_{0} $ from the origin\foot{The
restriction $y>-y_0$ is in fact imposed only on the trace, but it 
leads to a rapid  exponential decay of the wave function of
the fermions when $y<-y_0$.}, at 
temperature $\b$ (see Fig.6).    The wall 
should be sufficiently far, so that the  level of the Fermi sea is below the
top of the potential. The critical   distance   is
\eqn\ycrit{y_{0}^*=2 \sqrt{2(\pi -\b) N },}
which corresponds to  $T=T^*$.
Then the level of the Fermi sea reaches the top of the inverse gaussian potential.
The string  cosmological constant $\mu$,   which  measures the deviation from the critical point,  is   proportional to $(y_0^*-y_0)^2$.

 \epsfxsize=200pt
\hskip 20pt
\epsfbox{ 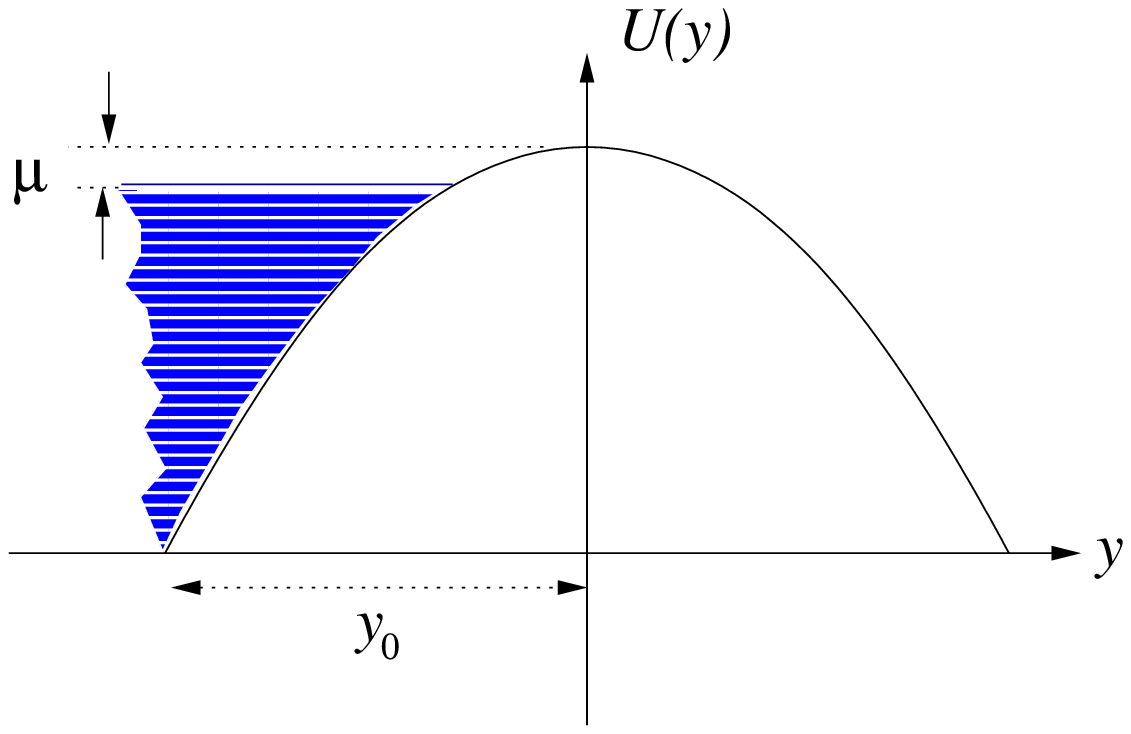   }
\vskip 5pt

\centerline{Fig.6 :    Fermi sea for  the inverse gaussian potential    }
 
 \vskip  10pt

 The system of $N$ fermions on a circle of length $\beta$ 
 describes the singlet sector of the
 Matrix Quantum Mechanics (MQM) \kleb\ and 
 gives a 
 non-perturbative
 realization  of the $c=1$ string theory compactified at length $\beta$.
 The partition function of this system, calculated in ref.
 \grkl , 
 is invariant  under the duality transformation 
 $${\b\over 2\pi} \to {2\pi\over \b}.$$
 The equivalence with the
 compactified at euclidean time interval $\b$ is consistent only
 in the interval $0<\b<\b_{KT}$, where $\b_{KT}=\pi$
 is the   Kosterlitz-Thouless point.
  Near  this point the the cutoff wall   approaches the top of the 
  inverse gaussian potential and  there is no  (metastable) ground state
  in the large $N$ limit.

  \appendix{B}{Fixing the parameters 
  of the solution}

  \subsec{The three conditions following from the asymptotics at
  $z\to\infty$}

Using the explicit form \SOLUTIONz\ of $z(u)$, we   rewrite   
\Jinfty\  as a series expansion in $u-\uy$:
  \eqn\ExpZETA{ \eqalign{
  \z(z)  = &\ { t_{2}e^{a+b} \over \sin \b} \left[ {H(2\uy )\over 
  H'(0)}\right]^{2} \ {1\over (\uy - u)^{2}}\cr
  &-\left[ {t_{1}
   e^{a+b\over 2}  \over 2\sin {\b\over 2}}{  H(2\uy 
  )\over
  H'(0)}
  +   2 {e^{a+b} t_{2} \over \sin \b}
  { H(2\uy ) H'(2\uy )\over  [H'(0)]^{2}} \right]\ 
  {1\over  \uy - u }\cr
  &+   C\cr
  &+ \CO( \uy - u  ),\cr}}
where 
\eqn\TATB{t_{1} = - {T\over 2\sin{\b\over 2}}, \ \ \ 
t_{2} = {T\over 2}
}
 and the constant $C$ is certain function of the
elliptic modulus, whose explicit 
form will not be needed.  
 This is to be compared with the first three terms in the 
  Laurent expansion of 
  \SOLUTIONZ
  \eqn\seriesZETA{\eqalign{
\z(u) = &\ 
{B  \over (\uy - u)^{2}}\cr
&+0\cr
&+ A + {1+k^{2}  \over 3} B\cr
 &+ \CO( \uy - u  ).\qquad\qquad\qquad
 \qquad\qquad\qquad\qquad\qquad\qquad 
  \cr}
}
 which gives
\eqn\CSTBz{\eqalign{ B 
 &=  \cot  (\b  / 2)\ {t_{1}^2 \over 8 t_2} 
 \ {H^2(2\uy)\over H'^2(2\uy)}\cr
   e^{a+b\over 2} 
& =  { \cot{\b\over 2}\over 2}  \ { H'(0)  \over
    H'(2\uy )}\cr
    A &=C- {1+k^{2}  \over 3} B.\cr}}

 \subsec{The normalization condition}
 The normalization \normden\ means that the 
 integral of $J(z)$ along the contour $\CC_{1 }Ä$ surrounding  its  upper cut 
 is equal to one. In terms of the variable $u$, this integral  takes the form
 \eqn\NORm{ {1\over 2\pi} \int_{-K+iK'} ^{K+iK'} du z'(u) \z(u) = 1.}
 %
Writing (using the standard notation $Z(u)= {\Theta'(u) \over   \Theta(u)}$)
$$ z'(y+iK')   = Z(\uy +y) +Z(\uy -y)$$
 $$\z(y+iK') = A + B k^2 \sn^2 (\uy - y),$$
 we arrive at the following integral
  \eqn\NCON{ \eqalign{  1 =& {1\over 2\pi} \int_{-K}^K dy [Z(\uy +y) +Z(\uy -y)]
 [A+Bk^{2}\sn^2   (u_\infty-   y)  ].\cr}}  
   With the help of the addition formula \AS , 17.4.35, we write it in the form
$$  1 =  {Bk^{2}\over 2\pi} [ Z(2 \uy)  I_{1} +  \sn (2\uy) I_{2}]
      $$
 where
 $$I_{1}= \int_{-K}^K dv  \ \sn^2  v  = {2(K-E)\over k^{2}}$$
 $$I_{2}= k^{2}\int_{-K}^K dv \ \sn(2\uy-v)\ \sn^3  v.$$
 The second integral can be decomposed  as
 $$  I_{2}=  {\cn\dn \over \sn} (2\uy) \left[ I_{1} -\tilde I_{2}\right]\ ,$$
 where
 $$\tilde I_{2}=  \int_{-K}^K dv  
   {   \sn^{2} v\over
 1- k^{2 }\sn^{2 } (2\uy) \sn^{2 }v }$$
 is a standard elliptic integral \BF , 435.02 :
 $$  \tilde I_{2}= 
{ 2K \ Z(2\uy) \over k^{2} \sn\cn\dn  (2 \uy)}.
 $$
   Substituting this in \NCON\ yields 
 \eqn\NORMZT{
 {2\pi\over B} = 2 (K-E)   [Z(2 \uy) +  \cn\dn(2 \uy)]- 2K {Z(2 \uy) 
 \over \sn(2 \uy)}.}

\appendix{C}{The special case $\beta =\pi$.}

  Here we check that the solution of the $O(2)$ matrix model 
  with quadratic potential is obtained  as the limit $\b\to\pi$ 
   (or
   $ \l\to 0$) of our general solution \SOLUTIONZ -\SOLUTIONz . 
   This limit is quite subtle and does not commute with the scaling limit 
   $q\to 0$.
    In this limit we have 
   \eqn\PARAMETERS{   T^{*}Ä= 32,\ 
     {T-T^{*}\over T} =  
{  k'^{2} K^{2}\over E^{2}} , \ \  e^{a}= {\pi \over 8E}, \ \ e^{b}Ä= k'e^{a}Ä. 
   }
    At   $\l =0$ we have more symmetry because the special points of 
    the map
    coincide with the quarter of the period,  $\uy =   u_{b} = K/2$, and
    the function  $e^{z(u)}Ä$  becomes 
   a Jacobi elliptic function 
 \eqn\zeDDD{e^{z(u)}Ä  =  - i  \  e^{a} \ 
 \dn\left( u- {K\over 2} -iK'\right) 
 = e^{a} Ä \  {\cn\over \sn} \left( u- {K\over 2} \right). }
  The   function $\z(u)$  is analytic in $z$ at $\l=0$
  %
   \eqn\zetaANl{ \z(u) = A +  {B\over \sn^2( u- K/2) }
   = A+B - B e^{-(a+b)} e^{2z(u)}}
 but the coefficient 
  $B$ tends to infinity as 
  $1/\l$ 
\eqn\BBb{  B= {\pi\over \l}{ T/T^{*}Ä\over E^{2}}Ä  .}
  Therefore, in order to take the limit $\l\to 0$, we should 
  first subtract  the diverging analytic piece of $\z$. The 
   finite, nonanalytic   piece  is then given by 
   \eqn\ZSING{\zeta (u) =\left(B\l {d\over  d\l}
   { 1\over \sn^2( u-  K/2 -\l K/2) } \right)_{\l=0} }
  where $${d\over  d\l} \equiv {\p \over \p \l }+ \left({\p u \over 
  \p\l}\right)_{\!\!z}
  { \p\over \p u} . $$  
     Further we write
 $$ \left({d\over  d\l} {1\over \sn^2( u- K/2 -\l K/2 ) }  \right)_{\l=0} 
 =   (K - 2(\p_\l u)_z ) \
 \left[ { \cn\dn   \over  \sn^3 }\right] \left(u-{K\over 2 }\right)  . $$
 $$K - 2(\p_\l u)_z   =2 K   \left[ {H'\over H}
 {\sn\cn \over \dn}\right]\left(u+{K\over 2}\right) 
     $$ 
   
   Since $     k'   \left[{\sn \over 
   \cn}  \right] \left(u+{K\over 2}\right) =-e^{z-a } 
   $, we can write $\zeta(u)$ as

  \eqn\ZEta{\eqalign{
  \zeta(u) & =   -  {KT     \over \pi}   e^{2z(u)}       
  {\p\over \p u} \ln  H  \left(u+{K\over 2}\right) .\cr
   }}
  The corresponding spectral density 
  $$\rho(z) =
  {  \zeta(u+iK')   +   \zeta(u-iK')  \over 2\pi} 
  $$
 coincides with the  one obtained by solving directly the saddle point 
 equations for the O(2) model
 \ref\MGaudin{M. Gaudin,  {\it unpublished notes} (1989).}.

 \listrefs

\bye